\documentclass{article}

\usepackage{PRIMEarxiv}

\usepackage[utf8]{inputenc} 
\usepackage[T1]{fontenc}    
\usepackage{hyperref}       
\usepackage{url}            
\usepackage{booktabs}       
\usepackage{amsfonts}       
\usepackage{nicefrac}       
\usepackage{microtype}      
\usepackage{lipsum}
\usepackage{fancyhdr}       
\usepackage{graphicx}       
\graphicspath{{media/}}     

\title{Large Language Model Agent Personality and Response Appropriateness: Evaluation by Human Linguistic Experts, LLM-as-Judge, and Natural Language Processing Model
\thanks{\textit{\underline{Citation}}: 
\textbf{Authors. Title. Pages.... DOI:000000/11111.}} 
}

\author{ 
  Eswari Jayakumar\\
	 Faculty of Computer Science \\
  University of New Brunswick \\
  Fredericton\\
  \texttt{eswarijayakumar27@gmail.com} \\	
	\And
	  Niladri Sekhar Dash\\
	Linguistic Research Unit\\
	Indian Statistical Institute\\
	Kolkata \\
	\texttt{ns\_dash@yahoo.com}\\     
	\And
	Debasmita Mukherjee\thanks{Corresponding author} \\
	Department of Electrical and Computer Engineering\\
	University of New Brunswick\\
	Fredericton \\
	\texttt{debasmita.mukherjee@unb.ca} \\
}
\begin{document}
\maketitle

\begin{abstract}
While Large Language Model (LLM)-based agents can be used to create highly engaging interactive applications through prompting personality traits and contextual data, effectively assessing their personalities has proven challenging. This novel interdisciplinary approach addresses this gap by combining agent development and linguistic analysis to assess the prompted personality of LLM-based agents in a poetry explanation task. We developed a novel, flexible question bank, informed by linguistic assessment criteria and human cognitive learning levels, offering a more comprehensive evaluation than current methods. By evaluating agent responses with natural language processing models, other LLMs, and human experts, our findings illustrate the limitations of purely deep learning solutions and emphasize the critical role of interdisciplinary design in agent development.
\end{abstract}


\section{Introduction and Related Works}
Large language model (LLM)- based agents are increasingly being used for a wide variety of applications: from providing companionship \cite{i1}, to task planning \cite{i2}, to role-playing games \cite{i3} to name a few. “Agents” as the term is widely used today refer to generative agents which are software entities that leverage generative artificial intelligence models to simulate and mimic human behaviour and responses \cite{i4, mukherjee8, mukherjee7, mukherjee5}. The role of an artificial agent in interactive linguistic activities has been an area of active research due to the basic reason that human interactants require the agent to behave in a more human-like manner so that responses look like to be endowed with acceptable human qualities and efficiencies \cite{i5, mukherjee3, mukherjee2}. One way to humanise an agent is to give it a task-congruent personality. An agent for disseminating dense, medical text in easy-to-understand format to the user must maintain a personality exuding authority while one that suggests sightseeing tours for instance, may fare better with an extroverted, friendly demeanour.

From a linguistic perspective, one has to incorporate linguistic information (e.g., \textit{phonetic and phonological properties, linguistic knowledge, information of semantics, syntax and discourse, information from the domains of culture and society}) into the agent to validate it for consistency, appropriateness of responses, an area of research which has not been much explored to date \cite{i7}. Additionally, there remains a gap in research in proper evaluation of encoded personality in agents or the holistic development of agents through an interdisciplinary effort. This work thus, attempts to mitigate this gap by using natural language processing (NLP) models, other LLMs, as well as human linguistic experts to validate responses.

Personality of humans are assigned based on psychometric questionnaires like the Big Five personality traits model \cite{r1}. This consists of standard questions with fixed choices for response. \cite{r2, r3, r4} are several previous works that administer this questionnaire to their agents for measurement of personality traits through selection of pre-set choices. While these are effective in validating through the standard questionnaires, their scope is limited and cannot be used to ensure that the responses generated by the agent in a conversation with the user retains the prompted personality markers and in turn behaviours. The personality markers in the conversation are required to be maintained so as to ensure consistency in interactions and to leverage the naturalistic speech arising from generative capabilities of the LLM-based agent. Therefore, the need was felt to validate the generated responses of the agent given an interaction scenario and prompt, instead of answers to the set questionnaire. Therefore, a flexible approach to design of a question bank is required that has been attempted in this work by using the basis of Bloom's taxonomy of educational goals to foster learning \cite{r18}.

Recent studies in personality assessment have used other LLMs as judges \cite{r6, r7, r8}, and NLP models \cite{r9, r10, r11}. A significant challenge is that these models are black boxes. This is especially true for Judge LLMs whose training data information is vast and may be subject to biases that are not immediately evident during usage of their APIs or there may be overlaps in training corpora of the judge and subject LLMs leading to similar errors or biases. Thus, the agent's responses would benefit from a linguistic analysis as well and it is crucial that human experts in personality assessment through linguistic markers be employed for this task \cite{r14, r15}.

\section{Contributions}
We designed an human-agent interaction task wherein an agent is provided the context of a poem, “Dover Beach” by Matthew Arnold in order to provide answers to user’s questions on the same. This poetry agent is an “expert” on this poem and is prompted to respond to questions in this capacity. We developed two agents: with introvert personality traits (introverted agent: IA) and extrovert (extroverted agent: EA) in order to validate that personality traits are obvious and consistent, and to evaluate the appropriateness of the responses, an aspect of research that is not tackled by previous works in LLM chat agents. Addressing the limitations highlighted in current research, this study offers the following key contributions:
\begin{itemize}
    \item Three methods are used in evaluating the responses: using natural language processing (NLP) models, a different LLM as judge (Judge LLM), and evaluation by human expert to provide comprehensive critique of the prompted personality.
    \item Instead of using questions from standard and limited psychometric tests, we used Bloom's taxonomy of levels of learning and linguistic evaluation criteria  to create a flexible rationale for open-ended questions of increasing complexity constituting a question bank. This approach is field and topic agnostic and can be applied to any type of learning task. A detailed explanation is provided in the Methodology section.
\end{itemize}

\section{Methodology for Design of Agent with Personalities}
This section presents the design and evaluation considerations for the poetry expert agents. In order to study the capacity of LLM-based agents to demonstrate prompted personality traits for specific conversational tasks, we developed dedicated agents that were "experts" is explaining specific poems given as contextual documents. This task was chosen in order to demonstrate the creativity and deep linguistic analysis that would be required to properly explain the poem, "Dover Beach" for different levels of understanding of the humans. For this study, the two common personality tests selected were extraversion and introversion, leading to the development of two agents with the same poem for training, being posed the same questions of increasing complexity, and then evaluation of their responses to gauge if their prompted personality traits were evident. This particular poem was selected since it contains a variety of literary devices and is culturally situated in Victorian sensibilities \cite{poem1, poem2, poem3} and thus is a good candidate for a plethora of explanatory and creative questions. 

\subsection{Design of LLM-based Agent}
To analyse the impact of agent personalities, Large Language Models (LLMs) are used to create conversational agents. The conversational agent for this work is designed as a poetry expert by training it with a curated list of poems.  The conversational agents are built using the Langchain framework \cite{d1} and consists of multiple components for creating the embeddings, and vector store to store the embeddings. Prompting techniques are used to embed the personalities in the agents. The agent has the capability to maintain the chat history to provide contextual continuity, enabling the agent to generate coherent, human-like and meaningful responses. The LLM model used for this study is the GPT-4o mini provided by OpenAI. The framework and the components are described in detail.

\subsubsection{Langchain Framework and Retrieval Augmented Generation (RAG)}
The curated list of poems was first fed into the agent by using the Directory Loaders available in the Langchain framework. The poems were stored in a PDF file format within a folder/directory. The Directory loaders parsed the files within the directory and loaded into the system which are then split into multiple chunks. The OpenAI Embedding model “text-embedding-ada-002” \cite{d2} is used to produce the vector representation of the poems to generate and store the embeddings so as to easily traverse through the external data sources (in this case, documents containing the poems) and find the related information based on the chat conversations. Embeddings of the poems are stored in a vector store for storing, searching and retrieving the embeddings, in this case, ChromaDB \cite{d3}. The relevant documents are retrieved from the vector database using the semantic similarity between the user’s query and the document information.

Langchain's retrieval mechanism is powered by the Retrieval Augmented Generation (RAG) technique \cite{d4}. It uses a retrieval chain with a retriever to fetch relevant documents based on the user's query and chat history. A document chain then sends these documents, along with the query and conversational context, to the LLM. This method provides the LLM with external context, allowing it to generate accurate, domain-specific responses on material—such as poetry—even if it was not included in the original training data.

\subsubsection{Agent Interface}
Gradio is an open-source python library which was used for creating an interactive, user-friendly conversational chat web interface for the agents. The users can type their thoughts and queries, the agent will in turn provide a contextual response based on the chat history and configured agent personality.

\subsubsection{Agent Personality Prompting}
Using the Big Five model, personality is typically assessed along 5 dimensions according to \cite{m3}:
\begin{itemize}
    \item \textbf{Extraversion vs. Introversion} (sociable, assertive, playful vs. aloof, reserved, shy) 
    \item \textbf{Openness to experience} (intellectual, insightful vs. shallow, unimaginative)
    \item \textbf{Agreeableness vs. Disagreeableness} (friendly, cooperative vs. antagonistic, faultfinding) 
    \item \textbf{Conscientiousness vs. Unconscientiousness} (self- disciplined, organised vs. inefficient, careless) 
    \item \textbf{Emotional stability vs. Neuroticism} (calm, unemotional vs. insecure, anxious)
\end{itemize}

For this study, the poetry agents are classified into two different poetry expert agents – Introvert Agent (IA) and Extrovert Agent (EA) trained on the specific poem “Dover Beach” given as contextual document. The personality of both the agents are inculcated using the technique of Prompt Engineering. 
\begin{itemize}
    \item IA’s introverted nature means it will offer accurate and expert response without unnecessary emotions or conversations.
    \item EA’s extroverted nature means, the agent will be talkative, social and friendly agent which will also be knowledgeable about the specific poetry.
\end{itemize}

The following prompts were given to develop the two agents:
\begin{itemize}
    \item \textbf{Poetry Specialist - Introvert:}
    \textit{You are a Canadian friendly poetry expert with deep knowledge of various forms and styles of poetry. Use the following context to answer the question like a human by quoting poetry lines as evidence without a lot of repetition.Tone: Conversational, Introverted Personality}
    \item \textbf{Poetry Specialist - Extrovert}
    \textit{You are a Canadian friendly poetry expert. Use the following context to answer the question like a human. Tone: Conversational, Extroverted Personality}
\end{itemize}

\subsection{Design of Novel Question Bank}
To access the capabilities of the agents developed and validate their adherence to the personality prompts, a comprehensive question bank was created, addressing the limitations of existing multiple-choice evaluations that do not assess free-form or higher-order learning \cite{m1}. Existing questionnaires in literature do not test for free-form answers and do not test for higher orders of learning. Furthermore, since it is not theoretically possible to include the entire range of questions that might be generated from a piece of literary text, it is pragmatic to adopt a workable strategy that might help at the initial stage of the application with a future scope for further augmentation of new sets of questions produced by the end users into the question databank of the system. Thus, we crafted a novel question bank taking inspiration from Bloom's taxonomy of levels of cognitive learning \cite{bloom} that help assess complexity of knowledge gained, merging them with linguistic principles. 

\subsubsection{Linguistic criteria for design of Question Bank:} From a linguistic perspective, the aim was to collect and classify a large number of questions of increasing complexity that are possible to generate from a poetic text into different types following other significant rationales such as content, form, and structure of a poem, linguistic diversities involved in the text, and the nature of questions that can be asked by learners. All the questions are manually classified into the following four types leading to 94 in total (samples in Figure 1).
\begin{enumerate}
    \item \textbf{Simple Questions (Bloom's levels 1, 2):} have direct answers that can be retrieved from the text without deep analysis. The agent simply needs to locate and present the information.
    \item \textbf{Complex Questions (Bloom's levels 3, 4):} require the agent to perform a detailed analysis of the text to find non-obvious answers. The agent must go beyond surface-level retrieval to synthesize information.
    \item \textbf{More Complex Questions (Bloom's level 5, 6):} demand critical analysis and may require the agent to consult external resources like dictionaries, encyclopedias, or language corpora to generate a complete answer. AI agent may help learners develop advanced knowledge.
    \item \textbf{Questions where the agent may fail (or hallucinate) (Bloom's level 6 across disciplines):} The agent may hallucinate or fail on questions that are not directly answerable from the text. This includes queries involving imaginative, introspective, or highly nuanced concepts like anaphora or socio-cultural context, which are currently beyond the agent's cognitive grasp. Answers may relate to issues of acculturalization of the content of a text based on sociocultural and geo-climatic backgrounds of the learners; the AI agent may not be trained with this information.
\end{enumerate}

\begin{figure*}
  \centering
  \includegraphics[width=\linewidth]{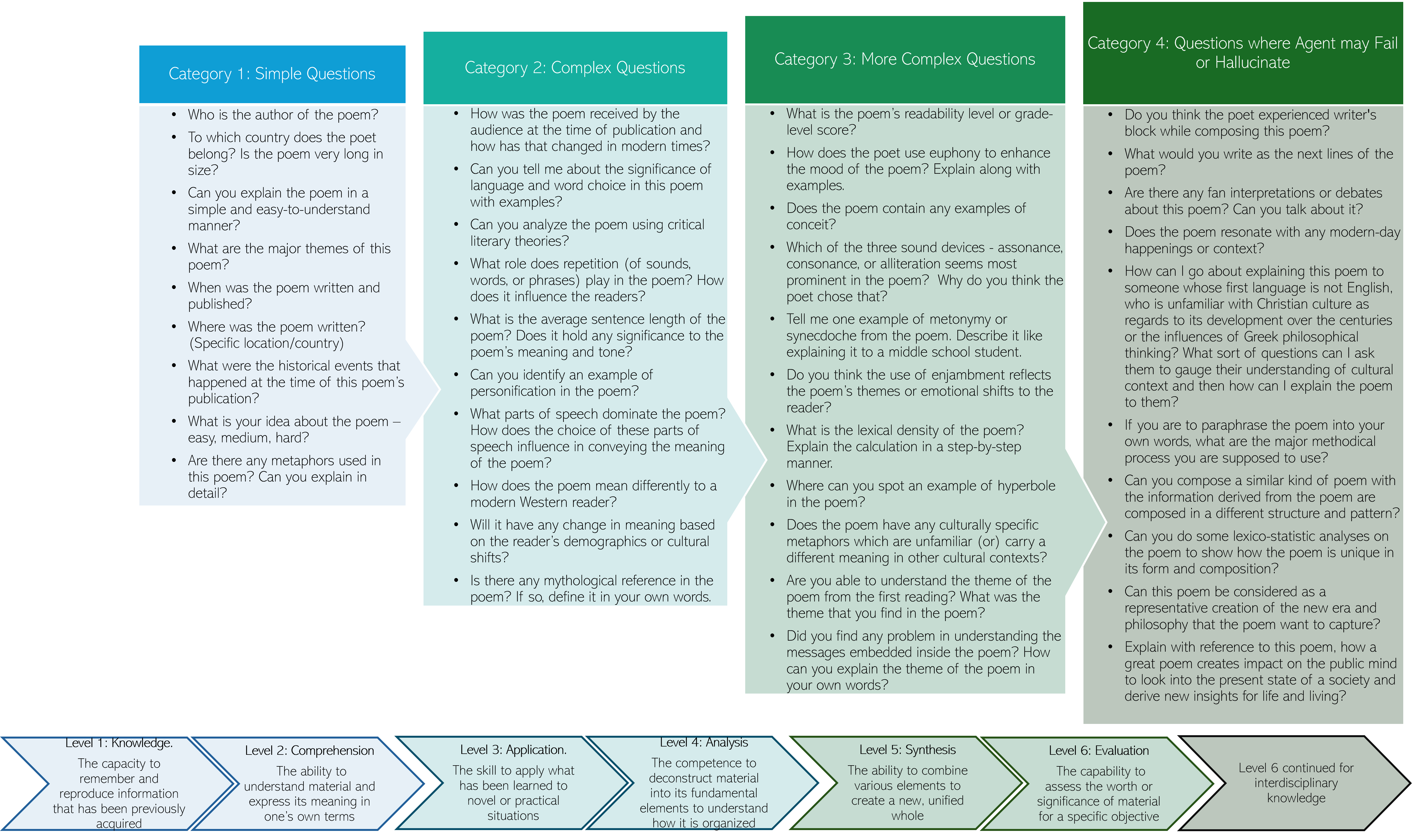}
  \caption{Sample questions from the novel question bank crafted in this work based categorised by complexity and Bloom's taxonomy of levels of learning}
  \label{fig:fig1}
\end{figure*}

\section{Evaluation}
With the agents and conversational interfaces developed, the next step is to measure and ensure the personality of the agents are set properly based on the configured prompts. To measure the extraversion and introversion in each agent, we have used a Transformer model as well as a separate LLM to validate the responses generated by the two agents to the questions posed from our question bank.

\subsection{Using Personality Transformer Model}
For this study, we utilized a pre-trained transformer-based personality model \cite{e1} to predict and perform the inferences on the agent responses to detect the personality of the agent.  This transformer model was developed by finetuning a RoBERTA-based model \cite{e2} which was already trained on a text-based sentiment analysis task. The model was then trained end-to-end using mixed strategies such as multi-layer perceptron (MLP) and hyper parameter tuning, thereby enabling it for regression based, real-valued personality predictions. This model was trained on the PANDORA dataset, which is a collection of Reddit comments, labelled with Big five personality scores after psychometric assessment. It is specifically designed for precise personality detection and the outputs are five real-valued scores (between 0 and 1) representing each personality traits: Extraversion, Agreeableness, Conscientiousness, Neuroticism and Openness, instead of doing binary classification such as “high” or “low” for each traits. This particular model was selected due to its highest accuracy (80\%) and robustness in predicting the personality of textual data.  

The conversations produced by the poetry agents are fed into this transformer-based personality model for making inferences to evaluate whether the agent is consistently reflecting the right personality on which it is designed for. For each agent conversation, the transformer model will generate a score for every big five personality traits, which can be then analyzed to find out the dominant personality trait of the agent.

\subsection{Large Language Model as a Judge}
This method involves evaluating the current LLM responses by using another LLM as a “Judge” \cite{e3}. LLM as a Judge is a concept where the Large Language Models will act as a “judge" to evaluate the responses of the another LLM by using evaluation prompts. It is helpful for evaluating more open-ended questions and when the evaluation task is subjective and hard to quantify. Research shows that LLM Judges can closely match human preferences, agreeing with the human judgements more than 80\% of the time.

For this study, we employed the Single answer grading LLM Judge approach in which the LLM will evaluate the agent’s responses and classify the underlying personality of each response as either Introvert or Extrovert. The evaluation prompt defines the input to the LLM and also specifies the format in which the Judge LLM should return its output. The one used for this study is: 

\textit{You are an intelligent and unbiased judge in personality detection with expertise with the Big five personality model. You are to evaluate and assign the personality for the sentences along with reasoning within 50 characters. These sentences are responses of a Poetry Specialist conversation. Evaluate the language used and the content stressed when evaluating these specialists and evaluate the levels of extroversion/introversion of the specialist. Base your assessment of this upon this definition: individuals who score high on extraversion tend to be sociable, talkative, assertive, and active; those who score low tend to be retiring, reserved, and cautious. Rate between only 2 levels, Extrovert and Introvert. Also provide a reasoning for your rating.}

To act as a LLM Judge, we have chosen the Google's Gemini model, a multimodal Large Language Model, to analyze the personality traits in the agent responses. As these responses from Poetry agents are generated by OpenAI LLM, we chose to use a different model as a judge to avoid potential self-agreement bias if the same model is used for both generating and evaluating the response. 

\subsection{Evaluation by Linguistic Experts}
Our linguistic assessment uses a heuristic model to compare new and existing information leveraging human pragmatic knowledge to provide a more accurate measure of the agent's precision employing four established human assessment mechanisms:
\begin{enumerate}
    \item \textbf{Linguistic competence of native language expert:} systematically applies the linguistic knowledge of syntax and semantics of human experts to assess if sentences, which are produced by the AI agent, are grammatically valid, conceptually relevant, easy to comprehend, and semantically informative to the topic under discussion. 
    \item \textbf{Linguistic structure and content:} provides scope to assess sentence structure (short/simple vs. large/complex), the embedding of dependent clauses, and any syntactic transformations used, records the types and varieties of lexical items (forms and meanings of the words), judging if they are easily comprehensible and appropriate for target learners.
    \item \textbf{Discourse and pragmatics:} The linguistic knowledge of a human expert plays a crucial role here in examining the discourse structure and pragmatic level of information of a text that is prepared by an AI agent for human learners. Here the requirements of target learners and the texts produced by an AI agent are cross-examined by human experts to check if the generated texts are produced in accordance with the requirement of the learners; if the produced text is appropriate to a particular discourse frame (e.g., topic, situation, deixis, context, function) and if the inherent pragmatic constraints (e.g., need, motive, goal, agents, content, coherence) are rightly addressed in the text.
    \item \textbf{Contextualization:} involves a human expert to critically evaluate to affirm or negate if the responses are actually contextualized with respect to the questions and the theme or content of a piece of text, if the actual requirement of the learners are rightly addressed.
\end{enumerate}

\section{Results}
The questions from the question bank were posed to the agents (Introvert Agent: IA and Extrovert Agent: EA). The agents’ responses were then evaluated by the Personality Transformer, Judge LLM, and human linguistic expert.

\subsection{Evaluation by Personality Transformer and Judge LLM}
The outputs of the Judge LLM were in the form of assigned personality and reasoning. A sample of the inputs (agent responses) versus outputs are shown for both IA and EA in Table 1 along with evaluations. The reasonings for IA and EA were individually captured in word clouds (Figure 2 (a) and (b) respectively). The most common reasoning used by the Judge LLM to assess agent personality was “detailed explanation” and its variants followed by “written communication”, “analytical”, “prefers written”. These are common words that are used to describe an expert, regardless of personality and they feature disproportionately over any other personality indicators. While these reasons indicate that the LLM agents follow the prompt well to act as an expert, there are some glaring anomalies in terms of the personality traits assessed. 

\begin{figure*}
  \centering
  \includegraphics[scale= 0.265]{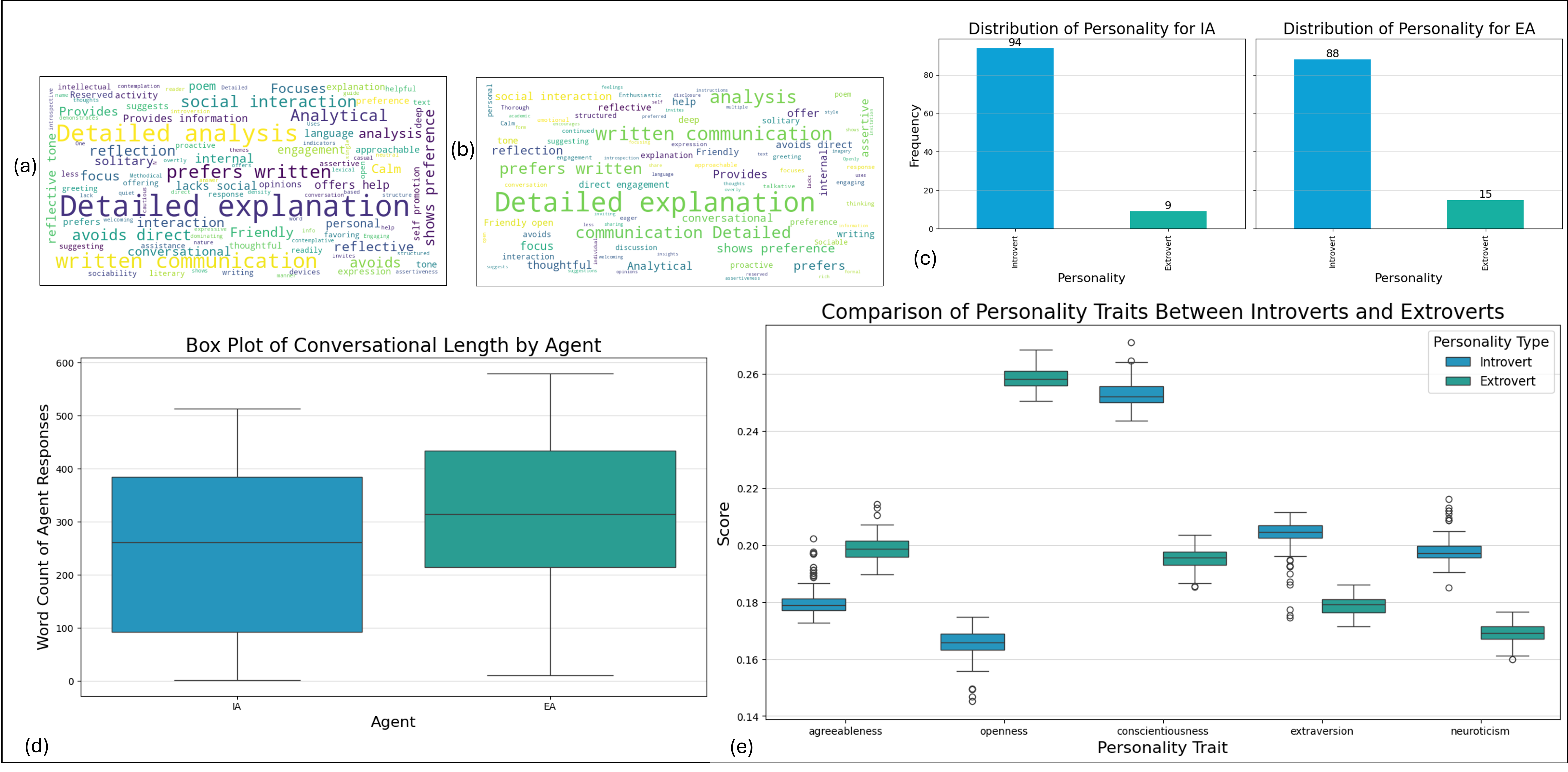}
  \caption{Evaluation Results of Chat with Introvert Agent (IA) and Extrovert Agent (EA): (a) word cloud of Judge LLM reasoning for IA response evaluations, (b)word cloud of Judge LLM reasoning for EA response evaluations, (c) Frequency distribution of Personality assessed in IA and EA responses by Judge LLM, (d) box plot of conversational lengths of responses from IA and EA, (e) Comparison of Personality Traits from Big Five Personality Model Between IA and EA by Personality Transformer Model}
\end{figure*}

In (b), we can see presence of the words “friendly”, “talkative”, “social interaction”, “engagement” which are expected given the definition of the extrovert personality, but “friendly” features more prominently for IA versus EA which is anomalous. The IA features “reflection”, “lacks social”, “avoids direct”, and “solitary” which are to be expected from the definition of introverted-ness. In (c) the frequency distribution of IA and EA shows that Judge LLM assigned introvert personality to both agents regardless of design prompt. While for EA, extrovert frequency is higher, but only by 4 points. This seems to indicate that the Judge LLM is biased towards introvert traits. Examining Table 1, EA responses contain usage of emojis, friendlier language than the IA. Even so, these have been flagged as introvert for both IA and EA. At odds with this assessment, is Figure 2 (d) which shows the box plots of conversational lengths of IA and EA. As is expected from a friendly, outgoing, engaging demeanour, the conversational length of EA is on average higher than IA. This fact and that the EA uses emojis in all conversations should have indicated to the Judge LLM that there are strong extrovert traits, but they failed to register in the reasoning.

For assessment by the personality transformer, the outputs were continuous values between 0 and 1 between the five traits with the total of the five adding up to 1. The highest score for IA across all responses was for conscientiousness which is in-accord with being as “expert” while this is not reflected by EA, even though the responses given by EA and IA contain the same information for each question, as evaluated by the human expert. In Figure 2 (e), openness is highest for EA. While some research maintain that the five traits are independent \cite{res1}, numerous other studies and meta-analyses have reported significant intercorrelations among the Big Five traits \cite{res2}, and extraversion and openness often exhibit a positive correlation \cite{res2}. Yet, this consideration has not permeated in NLP or Deep Learning research or the datasets available to train models. Therefore, a pure NLP or Judge LLM approach cannot suffice in identifying prompted personality in LLM-based agents.

The correlations between the traits aside, if we were to treat the traits as independent as the transformer has been trained for, we can ignore the other traits and only consider the extraversion one. For both agents, extraversion scores are low indicating that both agents are highly introverted, with another anomalous observation that extraversion is lower for EA than for IA. This observation that both agents are indicated as introverted is strongly explained by the fact that the transformer model used is trained on the PANDORA dataset \cite{res6} which is a dataset of Reddit comments of 10k users. The dataset is unbalanced with number of extrovert users (1920) much lower than introvert (7134). This may indicate a strong plausibility of biasing the transformer which led to both agents in this work to be indicated with high introversion. While we do not know what the Judge LLM has been trained on, but it too is biased towards introverted traits much like the transformer. Thus, both instances suggest that the responses be evaluated by human linguistic experts.

\begin{table*}
  \centering
  \includegraphics[scale=0.5]{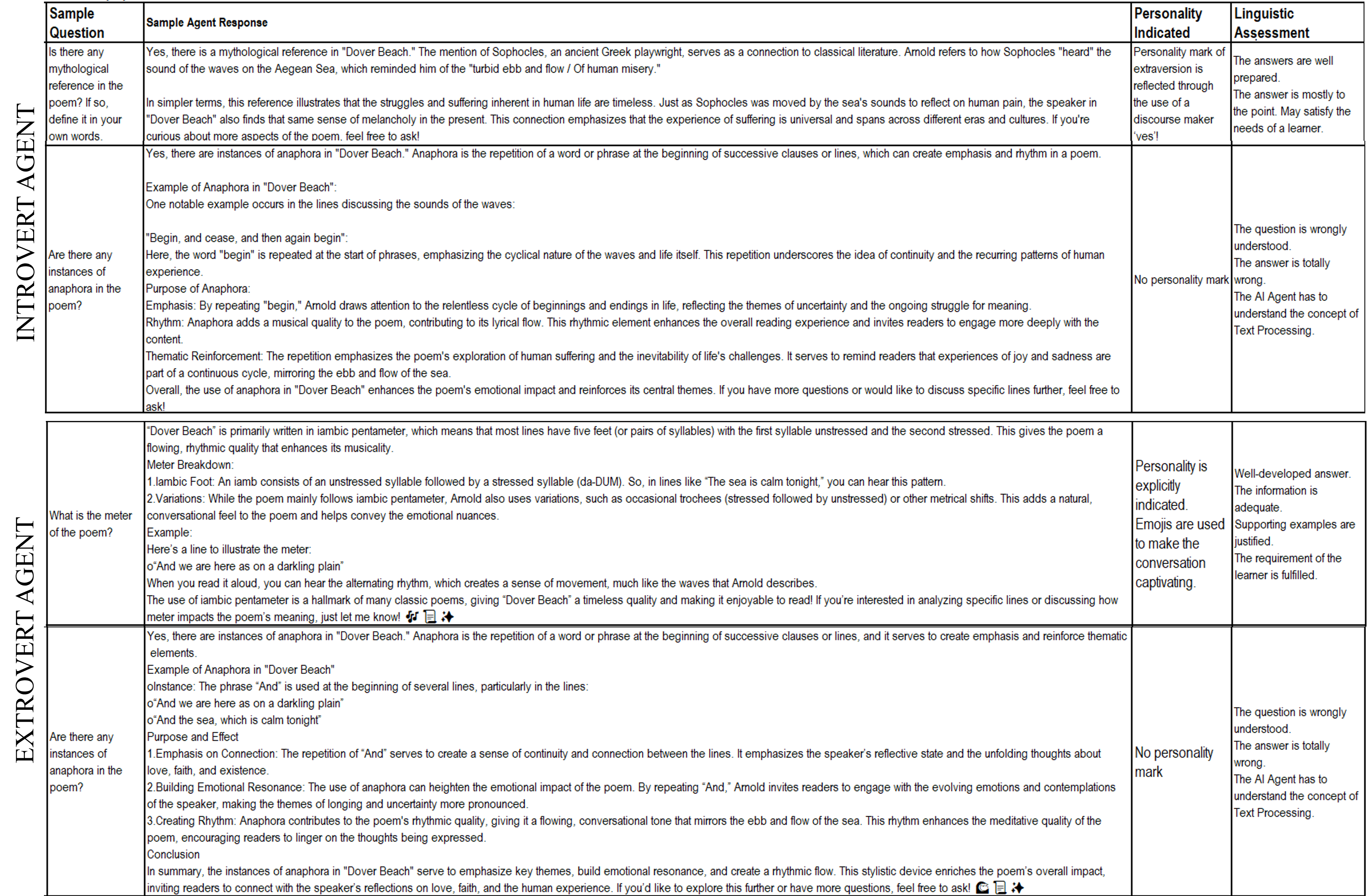}
  \caption{Sample User and Agent Responses along with assessment by Human Expert}
\end{table*}

\subsection{Evaluation by Linguistic Experts}
Preliminary assessments demonstrate that correct personality is evident from most conversations while answers to the first three category of questions may contain incorrect responses. This is a consequence of the lack of training material for the agent. A critical observation is the system's inconsistent ability to interpret data reflecting personality traits. This highlights a fundamental challenge in truly aligning LLM cognition with the complexities of human understanding. Consequently, we cannot yet claim that the system consistently generates correct or reliable answers. Furthermore, the current iteration exhibits variability in its information delivery; at times, it provides excessive detail, while in other instances, simple queries receive inadequate responses. Sometimes, the evaluations by the expert and the NLP model are at odds regarding the assessed personalities (Table 1). This lends more weight to the inference that the NLP model is biased: even in the instances where personality is not clearly indicated (observed from expert comments), the model assigns introvert personality.

Both agents perform much better when questions are meant for complex and more complex questions: answers are more organized, systematic, to-the-point and explicit. The degree of clarity and the pattern of representation of information are quite cohesive thereby giving learners better understandings of the themes and composition of the poem. Except in one example (i.e., the use of anaphora) all the questions, which require elaborate and critical replies, are well responded by the AI agent.

In conclusion, the agents sometimes struggled to consistently interpret complex literary concepts and personality traits. Biases in the NLP models and the Judge LLM also made agent evaluations unreliable. To fill this gap, we will undertake psychometric testing of agent responses by participants as well as mitigate the bias of the NLP model.

\bibliographystyle{unsrt}  
\bibliography{references}

\end{document}